\documentclass[twocolumn,showpacs,amsmath,amssymb]{revtex4}
\usepackage{graphicx}
\usepackage{color}
\begin{document}

\title{Qi-Wa, a problem that has plagued Chinese scrolls for millenniums}
\author{Ming-Han Chou$^1$, Wei-Chao Shen$^1$, Yi-Ping Wang$^1$, Sun-Hsin Hung$^2$, and Tzay-Ming Hong$^{1,3}$}
\affiliation{\\$^1$Department of Physics, National Tsing Hua University, Hsinchu 30013, Taiwan, Republic of China
\\$^2$Department of Registration and Conservation, National Palace Museum, Taipei 11143, Taiwan, Republic of China
\\$^3$Center for Fundamental Science Research, National Tsing Hua University, Hsinchu 30013, Taiwan, Republic of China}
\date{\today}
\begin{abstract}
Qi-Wa refers to the up curl on the lengths of handscrolls and hanging scrolls, which has troubled Chinese artisans and emperors for as long as the art of painting and calligraphy exists. This warp is unwelcomed not only for aesthetic reasons, but its potential damage to the fiber and ink. Although it is generally treated as a part of the cockling and curling due to moisture, consistency of paste, and defects from the mounting procedures, we demonstrate that the spontaneous extrinsic curvature incurred from the storage is in fact more essential to understanding and curing Qi-Wa. In contrast to the former factors whose effects are less predictable, the plastic deformation and strain distribution on a membrane are a well-defined mechanical problem. We study this phenomenon by experiments, theoretical models, and Molecular Dynamics Simulation, which give consistent scaling relations for the Qi-Wa height with length, width, thickness, and spontaneous curvature of the scroll. This knowledge enables us to propose modifications on the traditional mounting techniques, that are tested on real mounted paper to be effective at mitigating Qi-Wa. By experimenting on polymer-based films, we demonstrate the relevance of our study to the development of flexible electronic paper and computer screen. A new mechanism for plate tectonics is surmised that the variation of Poisson ratio at different depth can cause deformations similar to Qi-Wa on plate boundaries that are far away from the location where plates converge or transform. 

\end{abstract}
\pacs{46.32.+x, 46.70.De, 46.70.-p} 
\maketitle
	Chinese painting\cite{art1,art2,art3,art4} has a long history of tradition and styles in arranging human figures, animals, landscape, and plants into one or multiple themes as in symphonies. The collections of painting scrolls in National Palace Museum in Taipei, Taiwan can be dated back to the Six Dynasties (222-589) by Kaizhi Gu and Daozi Wu. 
The modes of landscape painting started in the Five Dynasties period (907-960) and have enjoyed many variations through the years. For instance, Kuan Fan, Xi Guo, and Tang Li of the Song Dynasty (960-1279) shifted their emphasis from grand mountains and waterfall to more intimate depictions as a result of the political and cultural shift to south China. Catering for the taste of emperors, painters also adjusted their focus on observing nature infused with poetic sentiments to expressing inner feelings. This focus on poetic sentiment led to the combination of painting, poetry, and calligraphy in the same work by the Southern Song (1127-1279).
Subsequent movements include the literati\cite{literati} painting in the Yuan Dynasty (1279-1368), concentration on personal cultivations and the establishment of separate local schools in the Ming (1368-1644) and Qing (1644-1911) dynasties. In the mean time, western painting techniques brought by the European missionaries have enthralled the Qing emperors and fused with the traditional ones to develop into an unorthodox yet popular style.

The image of hanging scroll paintings\cite{length} ranges from (1) medium square, e.g., 74.1 x 69.2 cm Traveling on a River After Snow, by Zhongshu Guo (?-977) of Song Dynasty, (2) medium long, e.g., 180 x 104 cm Wintry Forests, attributed to Cheng Li (916-967) also of Song Dynasty, to (3) oversize, e.g., 268 x 110 cm Pine Valley and a Cloudy Spring, attributed to Mao Sheng (ca. early 14th c.) of Yuan Dynasty. In contrast to hanging scrolls that are appreciated in its entirety while guiding the eyes through the artwork, handscrolls are intended to be viewed section by section during the unrolling and flat on a table. As a result, handrolls can be exceptionally long, measuring up to ten meters in the famous case of Along the River During the Ching-ming Festival, by Song artist Tse-duan Chang (1235-1305).

\begin{figure}[!h]
  \centering
  \includegraphics[width=8cm]{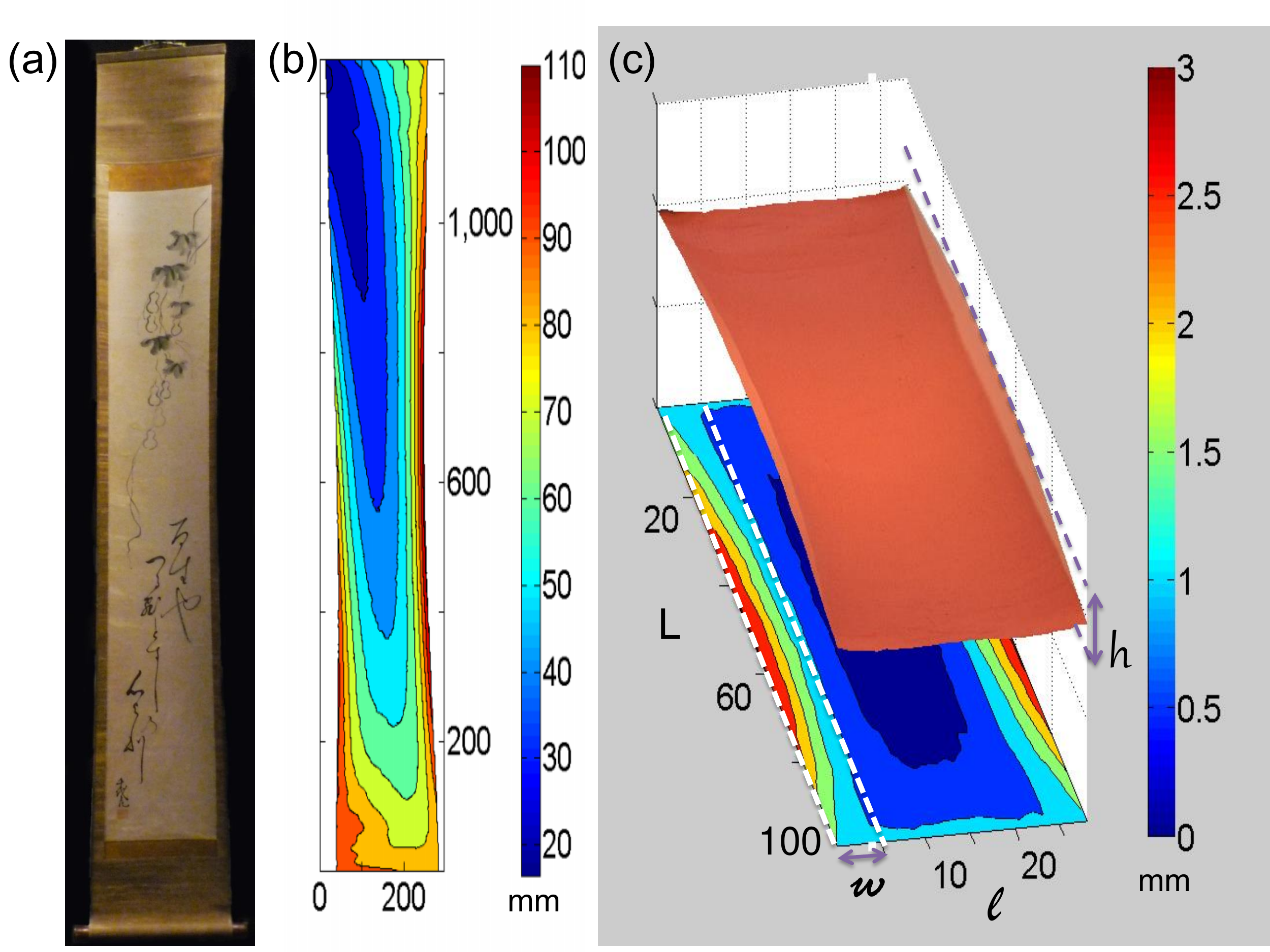} 
  \caption
  {(a) Qi-Wa is pronounced on the hanging scroll, One hundred gourds, all from the heart of a single vine, by Japanese painter Kaga no Chiyo (1703-1775). Image: 95.5 x 16.1 cm, and after mounting: 131 x 20 cm. Collection of Cornell Fine Art Museum. (b) and (c) are the scanned picture by a profilometer of (a) and PET film of $L=100$, $\ell=30$, and $1/\phi =10$ mm. Both exhibit U-shaped Qi-Wa.
}\label{exhibition} 
\end{figure}

The Chinese phrase, Qi-Wa (Qi means showing up and Wa refers to the roof tile), refers to the up curl on the sides of scrolls upon opening (see Fig.\ref{exhibition}). Trying to roll up any paper into a cylinder and give it a few strokes to consolidate the plasticity before spreading, one can easily witness and be convinced that this phenomenon is real and must have existed since the advent of scroll, be it Chinese or Egyptian. It not only presents an ugly appearance to the viewers, but may tear the fiber and cause pigment loss. Taping the midpoints of length to the wall will only make the matter worse because Qi-Wa now shifts to both one- and three-quarter-length positions. The reason why the phenomenon of Qi-Wa lingers for so long was perhaps partly due to the sectarianism in the craft of mounting\cite{ref}, where old masters were said to rarely exchange information to protect their own interest. The cultural and linguistic barrier also contributes to intimidating scientists from entering the branch of oriental art conservation and discovering the Qi-Wa problem.

Qi-Wa is generally included in the discussion of cockling and curling due to (1) quality of mounting paper\cite{daniels} or consistency on paste for the multiple backings\cite{mounting}, (2) residual tension from the ridding of creases and the lining and drying procedures\cite{nielsen}, (3) climate and conservation conditions such as humidity/moisture\cite{moisture,moisture2,wet} and temperature. Although the scroll is considered replaceable in Chinese tradition, any defect in scrolls can still lead to damages on the artwork since they are structurely joined together. The multiple components and materials (silk, paper, and wood) that compose scrolls make them a very complex object to predict. For instance, the warping due to different reactions to climate changes or detachment and crease formation from the action of rolling. Since scrolls are generally not intended for permanent display, they are rolled up and tied most of the time. Unlike the aforementioned factors which have an equal probability of generating a downward curl at the scroll boundary, the spontaneous extrinsic curvature $\phi$ incurred from the storage will be shown to be sufficient at causing and always favor an up curl, as is characteristic of Qi-Wa. This demonstrates that the spontaneous curvature is more essential to understanding and curing Qi-Wa. 

\begin{figure}[!h]
  \centering
  \includegraphics[width=8cm]{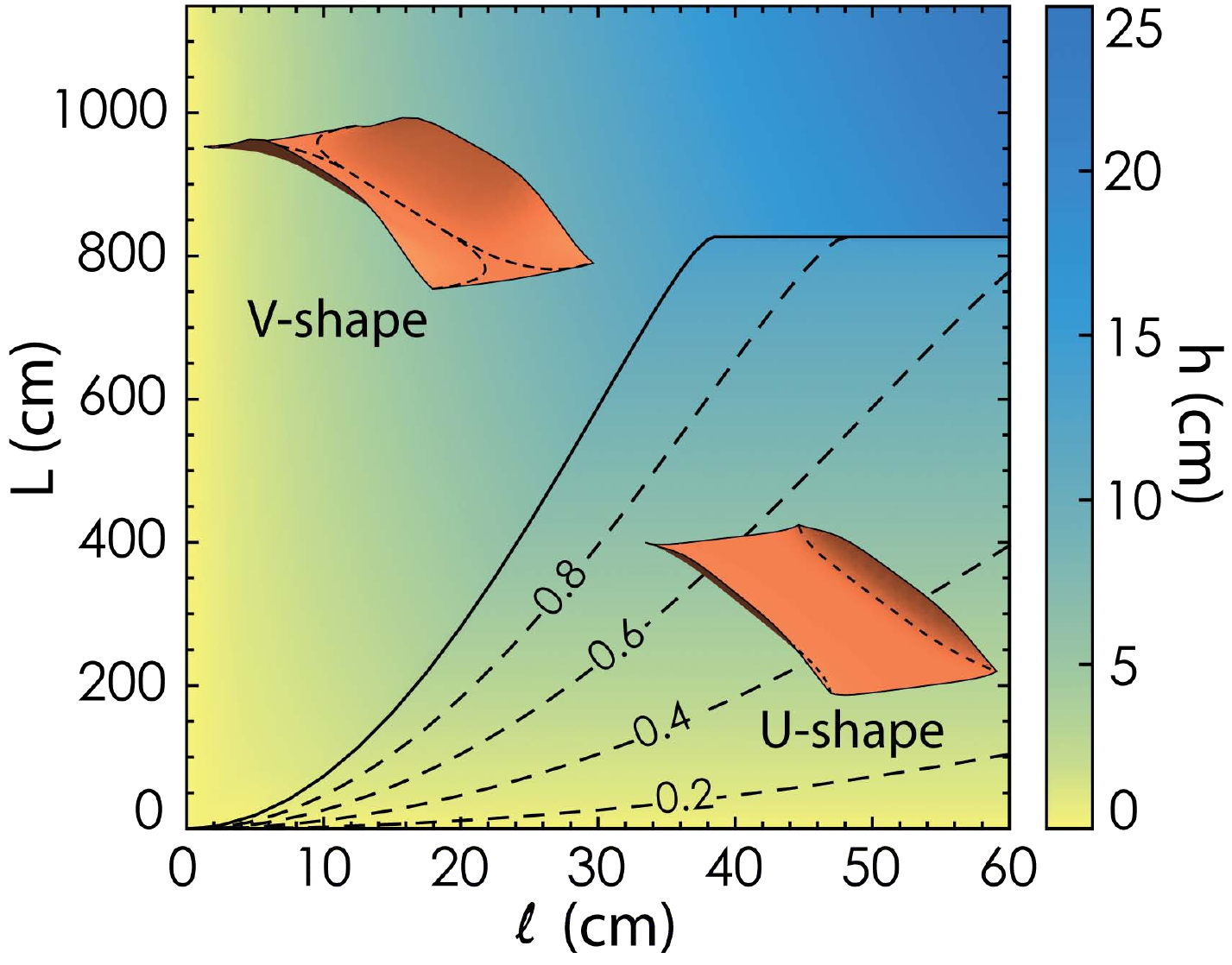} 
  \caption
  {Phase diagram for the U- and V-shaped Qi-Wa with $1/\phi =17$ mm, $\nu =0.2$, and $t=0.11$ mm. The contour plot illustrates the magnitude of $2w/\ell$, which equals 1 for the whole region of V shape. In the mean time, the density plot denotes the size of Qi-Wa height $h$. The phase boundary can be approximated by\cite{supplement} a parabolic function, $L\sim (\ell^2 /8) \sqrt{6\nu\phi/t}$, before leveling off at $L\sim 3.60/[(\nu\phi)^{1.5}\sqrt{t}]$. 
}\label{phase} 
\end{figure}

To focus on the role of spontaneous extrinsic curvature, we test on copy paper and Polyethylene terephthalate (PET) film instead of real mounted scrolls until the final stage of verifying our recipe to mitigate Qi-Wa. Papers are always stored in the dry box to minimize the effect of moisture. Besides being insensitive to moisture, PET film has the advantage of demonstrating the relevance of our study to the development of flexible electronic products. 
In most scrolls the deformation of Qi-Wa is confined to a narrow strip of width $w$ on both borders (see Fig.\ref{exhibition}(b,c)), which will be called U shape. In contrast, V shape is rare and occurs only when the width is so narrow that it allows the two Qi-Wa strips to touch.  As shown in Fig.\ref{phase}, these two phases\cite{leaf} are distinct with different physical origins and scalling relations. Other than this, no special attention will be paid in this work to distinguish between hanging scrolls and handscrolls because they behave similarly. 

First, let's explain the mechanism behind U-shaped Qi-Wa. While in storage, the outer layer of scrolls is plastically stretched as opposed to being compressed for the inner layer with a length difference, $\Delta L= Lt\phi$ where $L$ and $t$ denote the scroll length and thickness, and $1/\phi$ is the radius of curvature. After being unrolled, the bottom (originally the outer) layer gets squeezed, while the top one extended. This results in a transverse strain governed by the Poisson ratio $\nu$:
\begin{equation}
\frac{\Delta w}{w}=\frac{\Delta (\ell -2w)}{\ell -2w}=\nu \frac{\Delta L}{L}
\label{poisson}
\end{equation}
which equals $t$ times $\nu\phi$, the induced transverse curvature from the action of unrolling and a nonzero $\phi$. In Eq.(\ref{poisson}) we have purposely separated the strains on the Qi-Wa strips and the middle section. It is easy to imagine $\Delta w$ to cause Qi-Wa and write down $r\theta =w$ and $r(1-\cos\theta )=h$ where $r=1/(\nu\phi )$ and $\theta$ is the angular span of the warp. The latter relation reveals that $h$ has a maximum. In reality most warps are mild and $\theta$ can be treated as a small number to obtain:
\begin{equation}
w\approx \sqrt{{2h}/(\nu \phi )}.
\label{w}
\end{equation}
The up curl relieves the strain $\Delta w$, but pays the price of stretching Qi-Wa strips along the length by roughly $2h^2/L$ which raises the total energy by
\begin{equation}
\frac{K_{S1\parallel}}{2}\Big( \frac{2h^2 /L}{L}\Big) ^2 L\cdot 2w 
\label{seven}
\end{equation}
where $K_{S1\parallel}$ denotes the stretching modulus near the side borders (region 1) along the length.
In order to suppress this energy, the middle section (region 2) favors remaining flat and compressed with energy:
\begin{equation}
\frac{K_{S2\perp}}{2}\Big( \frac{\Delta (\ell -2w)}{\ell -2w}\Big) ^2 L(\ell -2w)=\frac{K_{S2\perp}}{2}(\nu\phi t )^2 L(\ell -2w)
\label{eight}
\end{equation}
where $K_{S2\perp}$ is the modulus perpendicular to the length. 
Essentially\cite{supplement}, it is the competition between Eqs.(\ref{seven}) and (\ref{eight}) that determines the optimal $w$ and $h$:
\begin{equation}
h\sim 
    \Big(\frac{K_{S2\perp}}{K_{S1\parallel}}\Big)^{1/4}L \sqrt{\frac{\nu\phi t}{6} }
  \label{Qiwa}
\end{equation}
where the lack of $\ell$-dependence is expected because Qi-Wa affects only a portion of the width. This scaling formula is not only of academic interest, but useful in practice. Namely, while $L$, $\nu$, and $\phi$ are hard to change, the mounting artisans are advised to weaken the $K_S$ along the transverse ($\perp$) direction in the middle section, while strengthening that along the longitudinal ($\parallel$) direction in the Qi-Wa strips to mitigate the U-shaped warp.

\begin{figure}[!h]
  \centering
  \includegraphics[width=8cm]{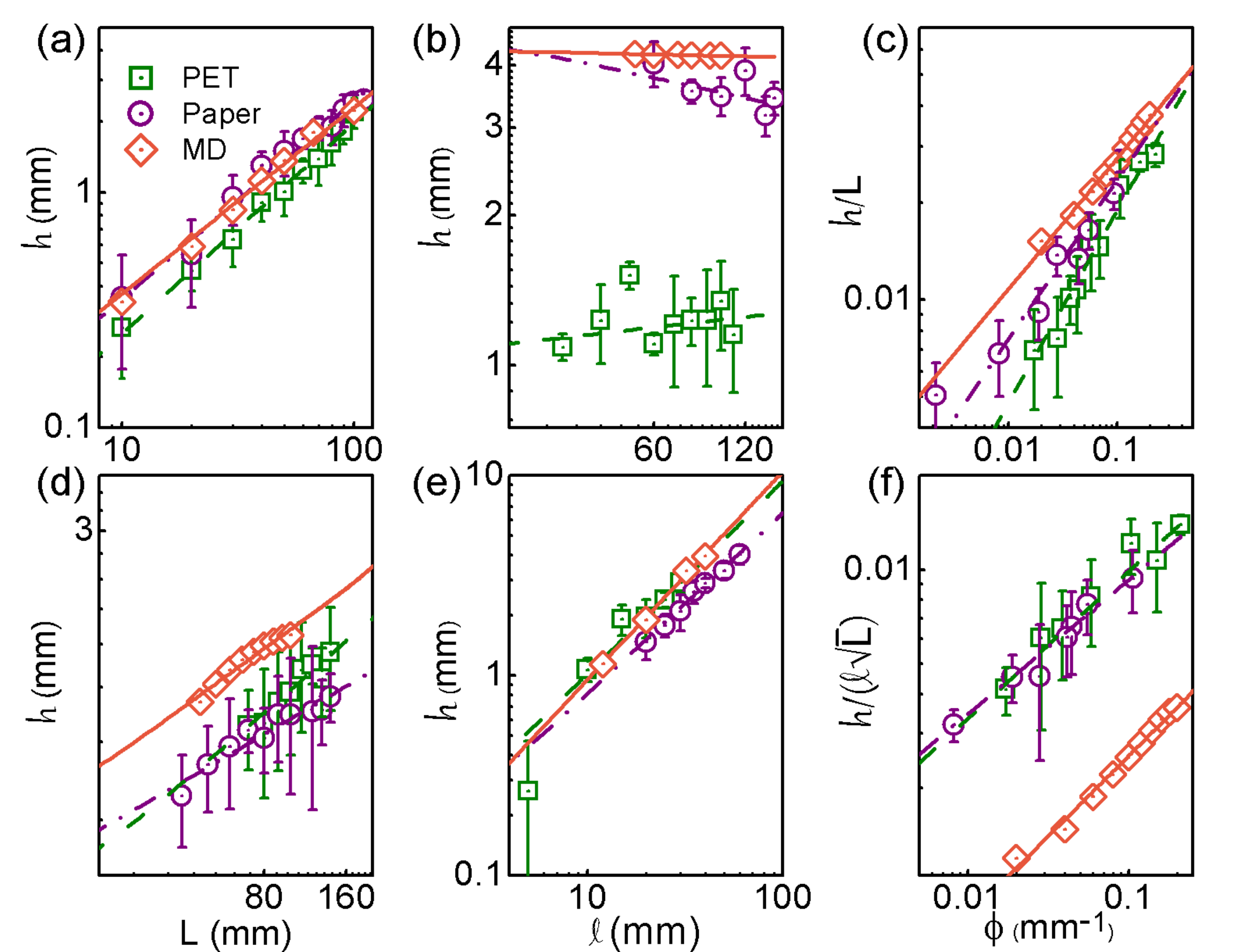} 
  \caption
  {Experimental and simulation results for the dependence of Qi-Wa height $h$ on the length $L$, width $\ell$, and spontaneous extrinsic curvature $\phi$. Straight lines are fittings by the scaling relation $h\sim L^\alpha \ell^\beta \phi^\gamma$ with exponents: (a) $\alpha$=0.9, 0.84, and 0.84 for PET film, copy paper, and simulation, respectively; (b) $\beta$= 0069, -0.133, and -0.013; (c) $\gamma$=0.6, 0.5, and 0.4 for U-shaped Qi-Wa.  In contrast, for V-shaped Qi-Wa  (d) $\alpha$=0.53, 0.36, and 0.45; (e) $\beta$=0.96, 0.9, and 1.06; (f)  $\gamma$=0.47, 0.43, and 0.51 The default size of PET film, copy paper, and simulated membrane are $(L,~\ell,~1/\phi)$= (60, 40, 10), (150, 60, 12.5) (all in mm), and (150, 80, 10) for U-shape data, as opposed to (120, 10, 6), (150, 20, 12.5), and (150, 40, 10) for V-shape. 
}\label{data} 
\end{figure}

Plugging Eq.(\ref{Qiwa}) to Eq.(\ref{w}) allows us to estimate how narrow a scroll needs to be in order to change from U- to V-shaped Qi-Wa:
\begin{equation}
\ell\le \sqrt{8L}\Big(\frac{t}{6\nu\phi}\Big)^{1/4}.
\label{boundary}
\end{equation}
The competing role of Eq.(\ref{eight}) is now played by the end segments (of width $w$) of the V, which straighten themselves to lower $h$ and lessen its pull on the Qi-Wa strips. The relevant energies are
\begin{equation}
\frac{K_{S1\perp}}{2}(\nu\phi t)^2 L\cdot 2w+\frac{K_{S2\parallel}}{2}\Big( \frac{2h^{\prime 2}}{L^2}\Big)^2 L(\ell -2w)
\label{vshape}
\end{equation}
where the down-graded stretching energy through modulus $K_{S1\parallel}$ has been neglected compared to the first term. The maximum height $h^\prime$ of the middle section that follows the induced curvature $\nu\phi$ obeys  $\ell/2 -w\approx \sqrt{{2h^\prime}/(\nu \phi )}$ similar to Eq.(\ref{w}) and $(h-h^\prime)/w=h^\prime/(\ell/2 -w)$ from similar triangles. By use of these conditions, Eq.(\ref{vshape}) can be minimized to give:
\begin{equation}
h\sim \Big(\frac{K_{S1\perp}}{K_{S2\parallel}} \Big)^{1/8}\frac{\sqrt{L}}{2^{7/4}}(\nu \phi )^{3/4}(t/3)^{1/4}\ell .
\label{vqiwa}
\end{equation}
Unlike Eq.(\ref{Qiwa}), $h$ becomes dependent on $\ell$, which is reasonable because $h$ must vanish as $\ell\rightarrow 0$, a limit only applicable to V shape. Another feature to note is the inverted ratio of $K_S$ and different exponent from Eq.(\ref{Qiwa}). As opposed to U shape, one needs to weaken the transverse modulus close to the border, while strengthening the longitudinal one in the middle section to suppress V-shaped Qi-Wa. 

\begin{figure}[!h]
  \centering
  \includegraphics[width=8cm]{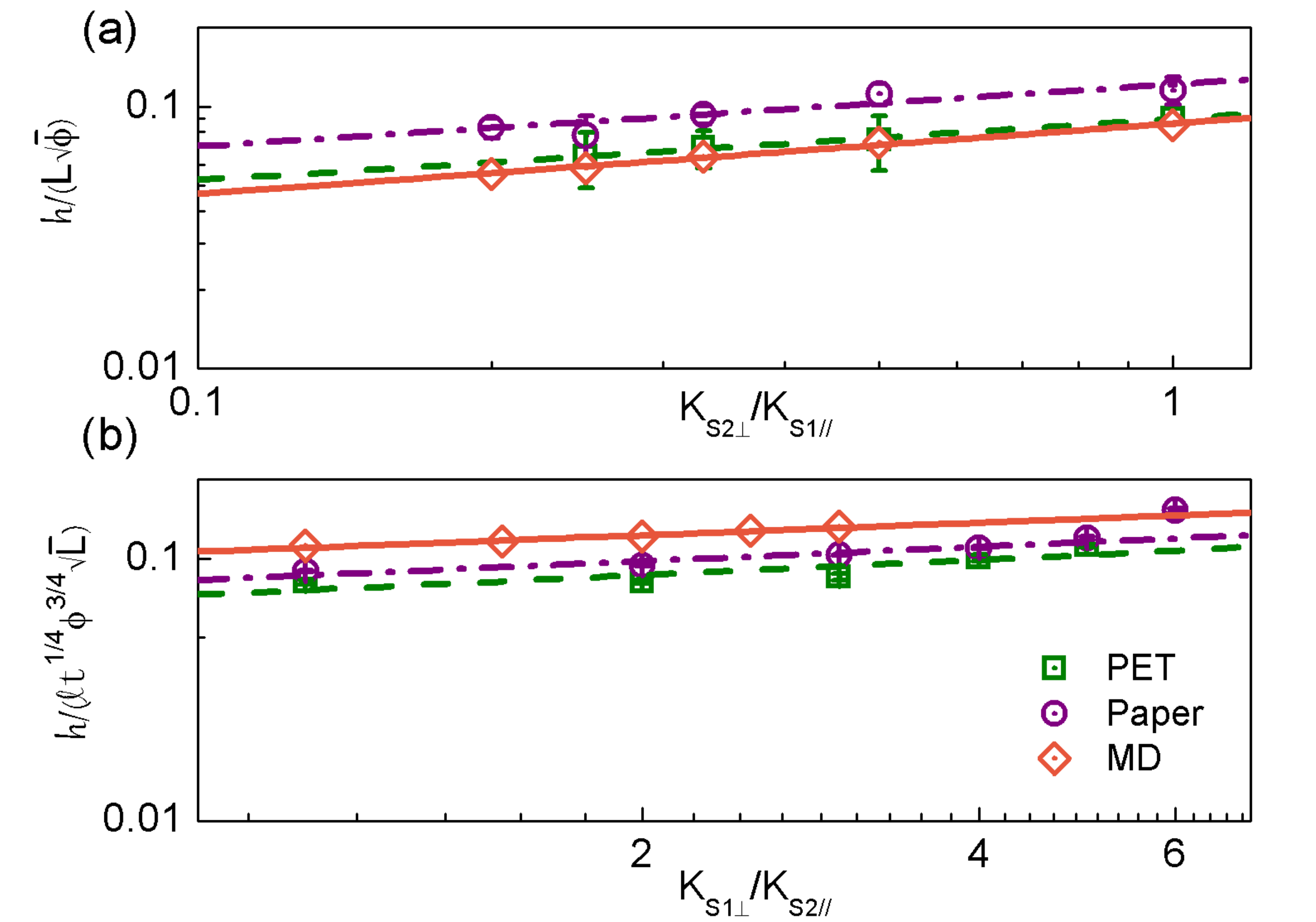} 
  \caption
  { (a) By adding layers to the U-shape Qi-Wa strips, the exponent of $K_{S}$ ratio in Eq.(\ref{Qiwa}) is measured to be 0.24 and 0.24 for PET ($L$=60, $\ell$=40, $1/\phi$=15, and $w$=10 mm) and copy paper ($L$=90, $\ell$=60, $1/\phi$=25, and $w$=15 mm). Simulation data are obtained by directly enhancing $K_{S}$ in the Qi-Wa strips ($L$=150, $\ell$=80, $1/\phi$=10, and $w$=20), which gives an exponent of 0.27. (b) Similar procedures determine the exponent of ${K_{S}}$ ratio in Eq.(\ref{vqiwa}) as 0.19, 0.18, and 0.16 for PET ($L$=80, $\ell$=15, and $w$=5 mm), copy paper ($L$=90, $\ell$=20, and $w$=5 mm), and simulation ($L$=150, $\ell$=40, $1/\phi$=10, and $w$=10) for V-shaped Qi-Wa. 
The value of curvature is not specified because it varies sensitively with the change of thickness, which effect is removed by renormalizing $h$ by $\phi^{3/4}$. 
}\label{kratio} 
\end{figure}

The different scaling relations in Eqs.(\ref{Qiwa}) and (\ref{vqiwa}) are verified in Fig.\ref{data} by experiments and Molecular Dynamics (MD) Simulation\cite{supplement}. A direct verification of their $K_S$ ratios, which provides a useful turning knob to minimize the warp, is achieved in Fig.\ref{kratio} by adding extra layers to region 1. This raises the $K_{S1\parallel}$ for U-shaped Qi-Wa in Eq.(\ref{Qiwa}) without affecting $t$ because the latter originates from Eq.(\ref{eight}) for region 2. Similar technique works for V-shaped Qi-Wa, but now the $t$ in Eq.(\ref{vqiwa}) belongs to region 1 and we need to renormalize $h$ by $t^{1/4}$ in Fig.\ref{kratio}(b) to extract the $1/8$ exponent of ${K_{S1\perp}}/{K_{S2\parallel}}$.

How did we realize that the spontaneous extrinsic curvature and Poisson ratio are crucial to explaining Qi-Wa? The first hint came from one visit to the National Palace Museum when we saw the curator placed pegs at every half meter on both lengths to stabilize the famous handscroll painting, Along the River During the Ching-ming Festival. The Qi-Wa is visible at all intervals and noticeably decreases as we moved towards the right. One would be hard pressed to ascribe this regular behavior to random mounting defects. We further noticed that the wooden roller used to roll up the scroll was attached on the left, which means the left section is originally in the inner core of the scroll. This helped us correlate the larger spontaneous extrinsic curvature with a more pronounced Qi-Wa. Second insight came from our initial failure in MD Simulations to observe Qi-Wa on a monolayer with build-in curvature. It was not until we added a second layer with a slightly longer bond length to simulate the effect of $\phi$ that Qi-Wa and all its scaling properties finally emerged\cite{simulation}. 

Other phenomena related to Qi-Wa can also be explained by our model: (1) Had we spreaded the sheet by pressing at the four corners, another pair of Qi-Wa would appear in the widths. They result directly from $\Delta L$ without invoking Eq.(\ref{poisson}) and so obey
\begin{equation}
h\sim 
    \Big(\frac{K_{S2\parallel}}{K_{S1\perp}}\Big)^{1/4}\ell \sqrt{\frac{t\phi}{6}}
    \end{equation}
similar to Eq.(\ref{Qiwa}) but with $K_{S\perp}$ and $K_{S\parallel}$ reversed, $L$ replaced by $\ell$, and $\nu$ deleted. (2) If we buckle the sheet into a convex arc, the Qi-Wa at $L/2$ will gradually diminish and reappear at both $L/4$ and $3L/4$ positions. (3) If the sheet is turned upside down, there is no Qi-Wa. (4) If the sheet is not tightly rolled so that $\phi$ becomes nonuniform, the Qi-Wa peak will shift towards the side that is originally in the inner circle where $\phi$ is the largest. 

\begin{figure}[!h]
  \centering
  \includegraphics[width=8cm]{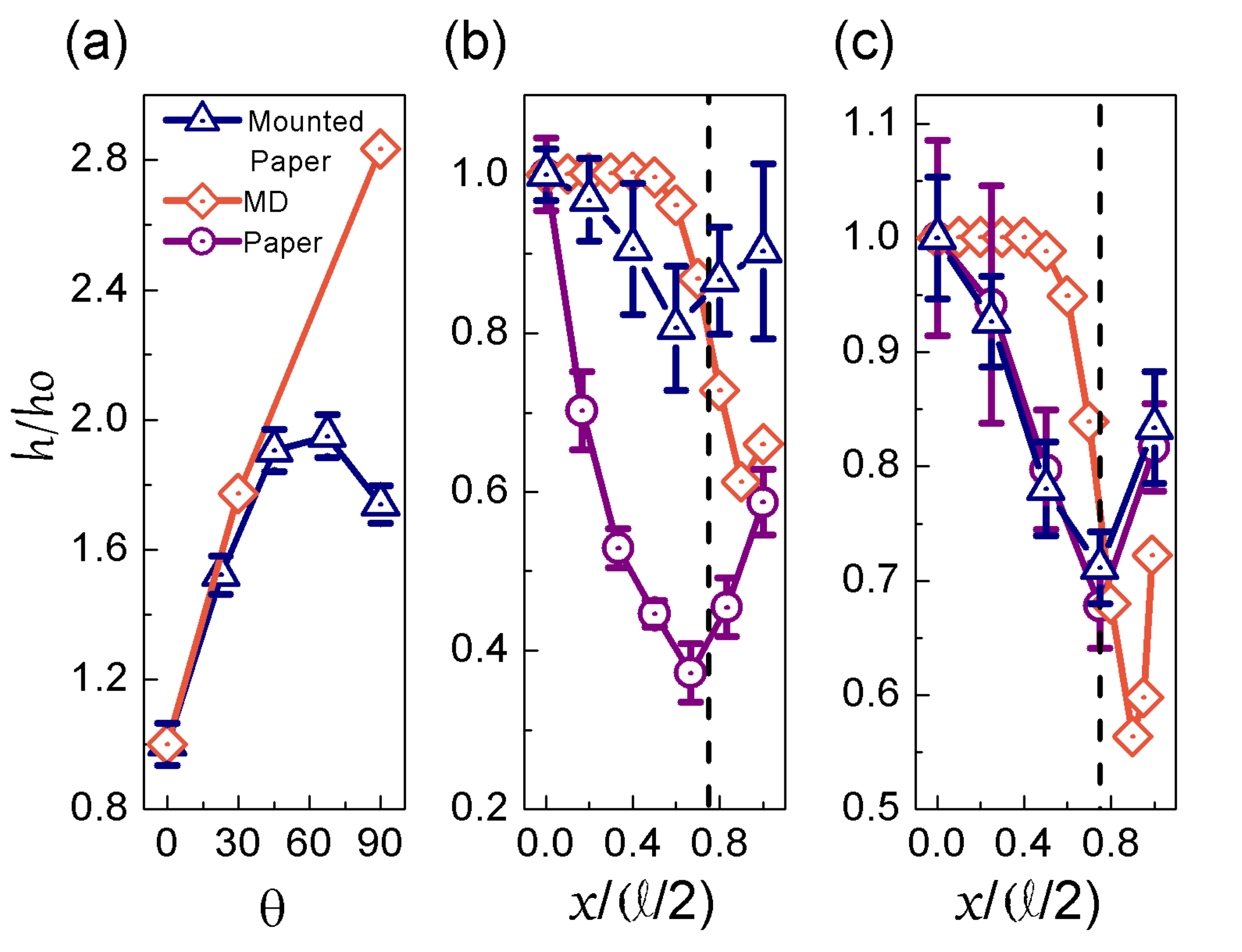} 
  \caption
  {Testing our proposals for mitigating U-shaped Qi-Wa. (a) Adjusting the relative angle $\theta$ between the length and main fiber orientation of mounting paper ($L$=100, $\ell$=55, $1/\phi$=20.2, and $t$=0.2 mm). Perfect alignment, $\theta =0$, gives the lowest $h$. (b) Adding an extra layer of width $(\ell/2) -x$ to the side borders of mounting paper ($1/\phi$=18.5 mm) and copy paper ($L$=110, $\ell$=60,$1/\phi$=16.5, and $t$=0.2 mm). The effect is optimal when the increase of thickness is limited to Qi-Wa strips. (c) Poking with a stiff brush on an area of width $2x$ centered on the backside of mounting paper ($1/\phi$=18.5 mm) and copy paper ($1/\phi$=18.6 mm). Poking in general suppresses $h$, but is most ideal to spare the Qi-Wa strips.
}\label{improvement} 
\end{figure}

Since mounting is an old trade with many traditions and strict aesthetic standards, any suggestion for mitigating Qi-Wa cannot deviate too much from the existing techniques. Our following recipe are intended for U-shape Qi-Wa and target at minimizing the $\nu^2{K_{S2\perp}}/{K_{S1\parallel}}$ factor in Eq.(\ref{Qiwa}). So far, our samples are limited to common copy paper and PET film, and it is yet to be vindicated that our mathematical equations and minimal model that does not include texture and paste is relevant to real scrolls. Obligatorily, we test our following proposals on real mounted papers in Fig.\ref{improvement}: (1) Fig.\ref{improvement}(a) shows that the most efficient remedy is to align the main fiber direction\cite{poisson} of backing papers which enjoys a larger $K_S$ along the length. This turns out to be opposite to the Chinese tradition, except for oversized scrolls. In contrast, the Japanese arrangement agrees with our suggestion, which may explain the general impression that Qi-Wa is less severe in Japanese and oversized Chinese scrolls. (2) Adding extra layers to the side borders helps because $K_{S1}$ is increased, as demonstrated by Figs.\ref{improvement}(b) and \ref{kratio}(a). This effect is maximized when the change of thickness is confined at region 1 affected by Qi-Wa. (3) Similar to the use of meat tenderizer in kitchens, poking on the backside of scroll with a stiff brush can weaken $K_S$ and decrease $\nu$. We are aware that this practice already exists in Japanese mounting procedures to enhance the adhesion of backing papers. However, instead of poking indiscrimately on the whole paper, Fig.\ref{improvement}(c) suggests to steer clear of region 1 to minimize Qi-Wa. (4) Minimizing the spontaneous extrinsic curvature by switching to a larger roller for storage. 

In conclusion, we show that the spontaneous extrinsic curvature incurred from storage is more essential to understanding and curing Qi-Wa than the conventional focus on moisture, paste, texture, and mounting skills. Like the latter factors, the strain distribution from the spontaneous curvature can be rigorously derived and modified by scientific means. Experiments, analytic derivations, and MD Simulations are employed together to find consistent scaling behavior for the Qi-Wa height. Although we strip off details like texture and paste in our minimal model, the remedies we propose to mitigate Qi-Wa are tested to be effective on real mounted paper. By sampling on PET films, we demonstrate
the relevance of our study to modern development of (1) polymer-based flexible electronic paper\cite{polymer}, (2) computer and cellphone screens that can be folded away or rolled up for storage\cite{fold,fold2}, and (3) any area-expanding devices of extraordinary expansion ratios since they require the use of either rolling-up or folding mechanisms of polymer films. For example, the solar cell screens\cite{solar} attached in satellites, the wings of air-borne objects, etc.. Plate tectonics\cite{tectonics} relates most of the deformation in the lithosphere to the interaction between plates, either directly or indirectly. Instead of concentrating on the area where plates converge or transform, we believe that warps similar to Qi-Wa will occur at plate boundaries that may be hundreads of miles away when the variation of Poisson ratio at different depths becomes significant.

We gratefully acknowledge funding support from National Science Council, technical supports by Hong Tan and Jofan Chien, and hospitalities of the Physics Division of National Center for Theoretical Sciences in Hsinchu and the National Palace Museum in Taipei, Taiwan. We also thank Itai Cohen, Xiang Chen and Yenchih Lin for allowing us to use Fig.1(a,b) and Pai-Yi Hsiao, Arnold C.-M. Yang, and Gerhard Gompper for helpful discussions.


\begin{thebibliography}{}
\bibitem{length} After being mounted, these numbers are increased by roughly one third.
\end{thebibliography}
\end{document}